\documentclass[aps,twocolumn,showpacs,floats,prb, eqsecnum]{revtex4}
\usepackage[german,english]{babel}
\usepackage{amsmath,amssymb}
\usepackage{graphicx}

\def\nn{\nonumber}

\def\be{\begin{equation}}
\def\ee{\end{equation}}
\def\bea{\begin{eqnarray}}
\def\eea{\end{eqnarray}}
\def\ba{\begin{array}}
\def\ea{\end{array}}

\newcommand{\bk}{{\bf k}}
\newcommand{\vfi}{{\bf {v}}_{F}^i}
\newcommand{\vdi}{{\bf {v}}_{\Delta}^i}

\begin{document}

\title{Signatures of the nematic ordering transitions in the thermal conductivity of $d$-wave superconductors}

\author{Lars Fritz}
\affiliation{Department of Physics, Harvard University, Cambridge MA 02138, USA}
\author{Subir Sachdev}
\affiliation{Department of Physics, Harvard University, Cambridge MA 02138, USA}

\date{\today}

\begin{abstract}

We study experimental signatures of the Ising nematic quantum phase transition in $d$-wave superconductors,
associated with the change of lattice symmetry from tetragonal to orthorhombic in the superconducting state.
 The characteristic feature of this transition is that the ratio between the Fermi velocity $v_F$ and gap velocity $v_\Delta$ flows to a maximally anisotropic fixed point, {\it i.e.} the renormalization group fixed point is situated at $\left ( v_\Delta/v_F \right)^*=0$. Our main point is that the logarithmic approach to this fixed point has visible signatures in the thermal transport. The analysis of the  quasiparticle contribution to the thermal transport is carried out in the framework of a kinetic approach, which shows
 that the thermal conductivity is enhanced near the nematic critical point. Another aspect of our study is the interplay of dilute disorder and electronic interactions  in the measured thermal transport coefficients.
\end{abstract}
\pacs{}

\maketitle


\maketitle

\section{Introduction}

Electronic nematic phases were discussed in the context of doped Mott insulators~\cite{Kivelson}, and have by now been experimentally observed in a number of systems. They have been observed in semiconductor heterostructures~\cite{Cooper}, in the bulk transition metal oxide $\rm{Sr}_3\rm{Ru}_2\rm{O}_7$~\cite{Borzi}, as well as in YBCO. In all these cases the thermal transition to the nematic phase seems to be second order, whereas the nature of the quantum phase transition is much less clear. In the case of $\rm{Sr}_3\rm{Ru}_2\rm{O}_7$ it seems to be first order~\cite{Borzi}. 

Now there is evidence that a nodal nematic phase occurs in at least some of the underdoped cuprate superconductors. In the nematic 
phase, the square lattice symmetry is broken down to tetragonal symmetry, a consequence of the instability
of the interacting electronic system to partial stripe-like order.
The best evidence for this comes from measurements of strongly temperature dependent transport anisotropies in underdoped ${\rm{YBa}}_2{\rm{Cu}}_3{\rm{O}}_{6+\delta}$~\cite{Ando} and from neutron scattering experiments in underdoped ${\rm{YBa}}_2{\rm{Cu}}_3{\rm{O}}_{6.45}$~\cite{Hinkov}. 

On the theoretical side, Vojta {\it et al.}~\cite{Matthias} analyzed possible quantum phase transitions in $d$-wave superconductors in the framework of a renormalization group (RG) analysis. An initial RG analysis found runaway flow for the nematic ordering instability at zero temperature based on an expansion in $3-d$, where $d$ denotes the spatial dimensionality. In a recent work, Kim {\it et al.}~\cite{Eun-Ah} found in the framework of a large-N analysis the existence of a second order transition to a nematic phase. Based on an RG analysis in the large-$N_f$ framework, where $N_f$ denotes the number of electronic spin components (the physical case corresponding to $N_f=2$), Huh~\cite{Yejin} {\it et al.} confirmed the existence of such a second order transition. The authors found an RG fixed point at order $1/N_f$ describing a second-order quantum phase transition associated with the onset of long-range nematic order. The scaling properties near the fixed point are very peculiar. There is a dangerously irrelevant parameter $v_\Delta/v_F$, where $v_\Delta$ and $v_F$ are the velocities of the nodal fermions parallel and perpendicular to the Fermi surface, which controls the fixed point. The fixed point lies at "infinite anisotropy", {\it i.e.} $(v_\Delta/v_F)^*=0$, which has to be contrasted from the other fixed points found for other competing orders, which are relativistically invariant~\cite{Matthias}. In order to calculate physical quantities one has to use a fully two-dimensional theory, since the flow of the anisotropy to zero is logarithmically slow as a function of the relevant energy scale, as is described by Huh {\it et al.}.\cite{Yejin} 

In the present paper we study thermal transport properties at the nematic to isotropic quantum phase transition (QPT) deep within the $d$-wave superconducting phase of a quasi two dimensional tetragonal crystal~\cite{Eun-Ah} in the framework of the Boltzmann equation. The main result of our analysis is the logarithmic enhancement of the thermal conductivity upon lowering the temperature, see Eq.~\eqref{thermalcon} and its disucssion in the result section together with Fig.~\ref{fig:vertex}.  

The paper is organized as follows. We start with a general review of the model for the second order nematic phase transition and its properties in a few limiting cases in Sec.~\ref{model}. This closely follows the presentation in Refs.~\cite{Eun-Ah,Yejin}. A review of the properties of the model under renormalization group transformations is given in Appendix~\ref{RG}. We proceed with the definition of the heat current operator of the electronic quasiparticles in Sec.~\ref{heatop} as we use it in the present work. In Sec.~\ref{Boltzmann} we introduce the Boltzmann equation framework and explain the variational ansatz used to solve it. In Sec.~\ref{disorder} we review a Boltzmann equation analysis of a disordered $d$-wave superconductor {\it without} a nematic mode, which makes connection to existing results~\cite{Graf,Durst}. Then we analyze the full problem taking into account elastic scattering from dilute impurities and inelastic scattering due to the nematic mode in Sec.~\ref{full}. The numerical solution of the Boltzmann equation is presented in the results section, Sec.~\ref{thermalfull} and several different situations are discussed, such as the interplay of inelastic and elastic scattering. Finally, in Sec.~\ref{conclusion} we conclude and comment on possible experimental implications of our analysis. Appendix~\ref{zeromode} provides some additional information on transport in clean systems and complements the discussion in Sec.~\ref{heatop}. Appendix~\ref{bosoniccurrent} further analyzes the problem in the framework of a Kubo formalism, and points out a couple of difficulties. This only serves to complement the picture and may serve as a starting point for more ambitious approaches to this complicated transport problem.

\section{The Model}
\label{model}
The model under consideration throughout this work has been discussed in the literature in great detail~\cite{Eun-Ah,Yejin}. Therefore we will only repeat the key features. The relevant low-energy description of the electronic system in a two-dimensional $d$-wave superconductor with a pure $d_{x^2-y^2}$ pairing symmetry is given by the following BCS-type Hamiltonian
\begin{eqnarray}\label{$d$-wave}
H &=&\sum_{\bf{k},\sigma} \Psi^\dagger_{1\sigma\bf{k}} \left (\begin{array} {cc} {\bf{v}}_{F}^1\cdot {\bf{k}} & {\bf{v}}_{\Delta}^1 \cdot {\bf{k}} \\ {\bf{v}}_{\Delta}^1 \cdot {\bf{k}} & -{\bf{v}}_{F}^1\cdot {\bf{k}}  \end{array} \right) \Psi^{\phantom{\dagger}}_{1\sigma {\bf{k}}} \nonumber \\ &+& \sum_{{\bf{k}},\sigma} \Psi^\dagger_{2\sigma{\bf{k}}} \left (\begin{array} {cc} {\bf{v}}_{F}^2\cdot {\bf{k}} & {\bf{v}}_{\Delta}^2 \cdot {\bf{k}} \\ {\bf{v}}_{\Delta}^2 \cdot {\bf{k}} & -{\bf{v}}_{F}^2\cdot {\bf{k}}  \end{array} \right) \Psi^{\phantom{\dagger}}_{2\sigma {\bf{k}}} \;,
\end{eqnarray}
where the dispersion has been linearized around the four nodal points.
The Fermi velocities ${\bf{v}}_{F}^1$, ${\bf{v}}_{F}^2$ and the gap velocities ${\bf{v}}_{\Delta}^2$, ${\bf{v}}_{\Delta}^2$ are defined as
\begin{eqnarray}\label{localbasis}
{\bf{v}}_{F}^1=\frac{\partial \epsilon_{\bf{k}}}{\partial {\bf{k}}} |_{{\bf{k}}={\bf{K}}_1} \quad  {\bf{v}}_{\Delta}^1=\frac{\partial \Delta_{\bf{k}}}{\partial {\bf{k}}} |_{{\bf{k}}={\bf{K}}_1}
\end{eqnarray}
and
\begin{eqnarray}
{\bf{v}}_{F}^2=-\overline{v} {\bf{v}}_{\Delta}^1 \quad {\bf{v}}_{\Delta}^2=\overline{v}^{-1} {\bf{v}}_{F}^1 \quad {\rm{with}}\quad \overline{v}=\frac{v_F}{v_\Delta} \;,
\end{eqnarray}
where the vectors ${\bf{K}}_i$ denote the location of the nodal points in the Brillouin zone in clock-wise direction starting with ${\bf{K}}_1$ lying at $(\pi/2,\pi/2)$. Furthermore, for later convenience, we introduced the anisotropy parameter $\overline{v}=v_F/v_\Delta$ in the above equations. This parameter plays a vital role since it is a direct measure of the velocity anisotropy and has a nontrivial flow under the renormalization group transformation~\cite{Yejin}. The main results of this RG analysis are subsummed in Appendix~\ref{RG}. Furthermore, it is notationally very convenient to introduce the nodal fermions, called $f_{1\sigma}$, $f_{2\sigma}$, $f_{3\sigma}$, and $f_{4\sigma}$ living at the respective nodes in k-space. The index $\sigma$ denotes a spin degree of freedom, {\it i.e.} $\sigma=\uparrow,\downarrow$. This can be extended to allow for a large-N treatment by generalizing $\sigma=1,...,N_f$. Our notation in the following will however stick to the SU(2) version. The Nambu spinors are composed of the nodal fermions in the following form 
\begin{eqnarray}
\Psi^{\phantom{\dagger}}_{1\sigma \bf{k}}=\left (\begin{array} {c}  f^{\phantom{\dagger}}_{1\sigma {\bf{k}}}  \\ \epsilon_{\sigma,-\sigma} f^\dagger_{3-\sigma -\bf{k}}   \end{array} \right ) \quad \Psi^{\phantom{\dagger}}_{2\sigma \bf{k}}=\left (\begin{array} {c}  f^{\phantom{\dagger}}_{2\sigma {\bf{k}}}  \\ \epsilon_{\sigma,-\sigma} f^\dagger_{4-\sigma -{\bf{k}}}   \end{array} \right ) \;,\nonumber \\
\end{eqnarray}
where $\epsilon_{\sigma,-\sigma}$ is the antisymmetric tensor.

The electronic Hamiltonian in Eq.~\eqref{$d$-wave} has a Dirac Hamiltonian structure and is readily diagonalized with a standard Bogoliubov transformation. This is achieved by a transformation according to
\begin{eqnarray}
\gamma^\dagger_{1\sigma \bk}&=& u_{\bk} f_{1\sigma \bk}^\dagger-v_{\bk} f_{3-\sigma-\bk}^{\phantom{\dagger}} \;, \nonumber \\ \gamma^{\phantom{\dagger}}_{3-\sigma -\bk}&=& u_{\bk} f_{3-\sigma-\bk}^{\phantom{\dagger}} +v_{\bk} f_{1\sigma \bk}^\dagger\;,
\end{eqnarray}
and respectively for nodes $2$ and $4$; $u_{\bk}$ and $v_{\bk}$ are the coherence factors~\cite{Schrieffer}. The corresponding unitary matrix is given by
\begin{eqnarray}\label{unitary}
U_{a\mathbf{k}}^{-1}=\frac{1}{2\epsilon_a({\bf k})}\left(
\begin{array}{cc}
{\bf v}_\Delta^a {\bf k} & -{\bf v}_\Delta^a {\bf k} \\
\epsilon_a({\bf k})-{\bf v}_F^a {\bf k} &\epsilon_a({\bf k})+{\bf v}_F^a {\bf k}
\end{array}%
\right) \;.
\end{eqnarray}
where the energies for the Bogoliubov quasiparticles are given by
\begin{eqnarray}\label{energies}
\epsilon_a({\bf{k}})= \left \{ \begin{array} {c}  \sqrt{\left( {\bf{v}}_{F}^1\cdot{\bf{k}}\right)^2+\left( {\bf{v}}_{\Delta}^1\cdot{\bf{k}}\right)^2}  \quad a=1,3   \\  \sqrt{\overline{v}^2\left( {\bf{v}}_{\Delta}^1\cdot{\bf{k}}\right)^2+\overline{v}^{-2}\left( {\bf{v}}_{F}^1\cdot{\bf{k}}\right)^2}  \quad a=2,4  \end{array}  \right. 
\end{eqnarray}
In our later discussion of the thermal transport properties in terms of the Boltzmann equation it is mandatory to work in the basis of these quasiparticles~\cite{FSMS,MFS}.

In the theory considered~\cite{Eun-Ah,Yejin,Matthias} an Ising-symmetric nematic order parameter couples to the nodal fermions. The corresponding interaction term has the form of an additional s-wave order parameter, whose condensation has the effect of breaking the four-fold rotation symmetry in k-space down to a two-fold one~\cite{Matthias,Eun-Ah}. 
Consequently, it assumes the following form
\begin{eqnarray}\label{eq:inter}
\mathcal{S}_{\rm{int}}=\lambda  \sum_{\sigma=1}^{N_f}\int d^2x d\tau \phi\left (\Psi^\dagger_{1\sigma}\tau^x\Psi^{\phantom{\dagger}}_{1\sigma}+ \Psi^\dagger_{2\sigma}\tau^x\Psi^{\phantom{\dagger}}_{2\sigma} \right) \;,
\end{eqnarray}
where $\phi$ denotes the Ising-type order parameter, whose action can be obtained by integrating out the fermions.
Kim {\it et al.}~\cite{Eun-Ah} showed that the leading quantum fluctuations at the large-N level after a quadratic expansion around the saddle point lead to a non-analytic form of the effective bosonic theory. This calculation is straightforward but tedious~\cite{Fradkin,Auerbach}, relying on an application of Feynman parameters. Retaining only the terms which are relevant at low energies in the RG sense (assuming the mass of the bosonic action is tuned to criticality), the effective bosonic action assumes the following form
\begin{eqnarray}\label{eq:bos}
\mathcal{S}_{\phi}&=&\frac{\gamma}{2 \beta} \sum_n \int \frac{d^2 k}{(2\pi)^2}  \left( \frac{\omega_n^2+\epsilon_1^2({\bf{p}})-\left( {\bf{v}}_{\Delta}^1 {\bf{p}}  \right)^2}{\sqrt{\omega_n^2+\epsilon_1^2({\bf{p}})}}\right. \nonumber \\ &+& \left. \frac{\omega_n^2+\epsilon_2^2({\bf{p}})-\overline{v}^{-2}\left( {\bf{v}}_{F}^1\cdot{\bf{p}}\right)^2}{\sqrt{\omega_n^2+\epsilon_2^2({\bf{p}})}} \right)\phi({\bf{p}})\phi(-{\bf{p}})\; ,
\end{eqnarray}
where 
\begin{eqnarray}
\gamma=\frac{\lambda^2}{32 v _F v_\Delta}\;.
\end{eqnarray}
This implies that the effective propagator of the bosonic modes is given by
\begin{eqnarray}\label{boseprop}
D^{\omega_n}_{\bk}&=&\frac{1}{\gamma}\left (\frac{\omega_n^2+\epsilon_1^2({\bf{p}})-\left( {\bf{v}}_{F}^1 {\bf{p}}  \right)^2}{\sqrt{\omega_n^2+\epsilon_1^2({\bf{p}})}} \right. \nonumber \\ &+&  \left. \frac{\omega_n^2+\epsilon_2^2({\bf{p}})-\overline{v}^2\left( {\bf{v}}_{\Delta}^1\cdot{\bf{p}}\right)^2}{\sqrt{\omega_n^2+\epsilon_2^2({\bf{p}})}} \right)^{-1} \;.\end{eqnarray}
Higher order terms are as usual in large-N theories down in powers of $1/N_f$. 
At this point it is instructive to discuss the above propagator in two extreme cases, namely for $\overline{v}=1$, {\it i.e.} isotropic velocities, and for $\overline{v}\gg 1$.

For $\overline{v}=1$ the corresponding bosonic propagator is simply given by
\begin{eqnarray}\label{isotropic}
D^{\omega_n}_{\bk}&=&\frac{1}{\gamma}\left (\frac{2 \omega_n^2+\epsilon^2({\bf{p}})}{\sqrt{\omega_n^2+\epsilon^2({\bf{p}})}} \right)^{-1}\;,
\end{eqnarray}
where the energies $\epsilon_1$ and $\epsilon_2$ are now identical due to the perfect isotropy. It is important to note that the present propagator implies the existence of quasiparticle-like peaks sitting on top of a continuum in the spectral function~\cite{Eun-Ah}.
The infrared RG fixed point of the theory lies at infinite anisotropy. In the limit of extreme anisotropy, {\it i.e.} $\overline{v} \gg 1$, the propagator reads
\begin{eqnarray}\label{aniprop}
D^{\omega_n}_{\bk}=\frac{1}{\gamma} \left (\sqrt{\omega_n^2+({\bf v}_F^1\cdot{\bf{p}})^2} + \sqrt{\omega_n^2+({\bf v}_F^2\cdot{\bf{p}})^2 }\right)^{-1}\;.
\end{eqnarray}
It was pointed out before that the presence of the nematic order parameter and its interaction with the Bogoliubov quasiparticles leads to a finite lifetime for the nodal quasiparticles~\cite{Eun-Ah}. 

In order to discuss transport in a realistic setting we will also add some dilute disorder. On a Hamiltonian level this reads\begin{eqnarray}
H_{\rm{dis}}=\int \sum_{i,j,\sigma}\frac{d^2 k}{(2\pi)^2} \frac{d^2 k'}{(2\pi)^2} V^{ij}_{{\bf{k}}{\bf{k}}'} f^\dagger_{i\sigma{\bf{k}}} f^{\phantom{\dagger}}_{j\sigma{\bf{k}}'} \;,
\end{eqnarray}
where $V^{ij}_{{\bf{k}}{\bf{k}}'}$ stands for the scattering matrix element of an electron living around node $i$ with momentum ${\bf k}$ into an electron living around node $j$ with momentum ${\bf k}'$. In contrast to Refs.~\onlinecite{Durst,dellanna} we assume a source of isotropic scattering, which implies that $V^{ij}_{{\bf{k}}{\bf{k}}'}=V^{\phantom{ij}}_{{\bf{k}}{\bf{k}}'}=\hat{u}$. However, all the results of the following calculations can in principle be extended to account for the more generic case including different scattering strenghts. As a remark it is important to note that our results will not recover the unitary limit.

\section{The kinetic approach to heat conductivity}

\subsection{The quasiparticle expression for the heat current}\label{heatop}

In our present paper we will concentrate on the contribution carried solely by the nodal quasiparticles. The heat current carried by the Bogoliubov quasiparticles is readily given by the intuitive expression
\begin{eqnarray}\label{heatcurrent}
{\bf j}_E= \sum_{\sigma=1}^{N_f}\sum_{i=1}^4 \int \frac{d^2 {\bf k}}{(2\pi)^2} \epsilon_i ({\bf k}) \frac{\partial \epsilon_i({\bf k})}{\partial {\bf k}} f^i_\sigma({\bf k}) \; .
\end{eqnarray} 
This expression will constitute the starting point for our Boltzmann transport equation analysis in Sec.~\ref{Boltzmann}.

For a general superconductor, irrespective of the pairing symmetry, we can refer the reader to Refs.~\cite{Moreno,Graf,Durst,Paaske,Durst2} for a proper derivation of the appropriate operator.

For an outline of a generic expression for the heat current in an interacting system we have added some remarks in Appendix~\ref{bosoniccurrent}, which also addresses the question of the role of the bosonic fluctuations.

\subsection{The Boltzmann equation}\label{Boltzmann}

Within this section we access the thermal transport properties using the semiclassical Boltzmann-equation approach. This approach has proven to be a powerful tool to compute transport properties of quantum critical systems~\cite{SSbook,Damle,SS,SVS,FSMS,MFS}.
In the following we assume that a quasiparticle description applies~\cite{Eun-Ah}. The controlling parameter of the Born approximation is given by $1/N_f$.

The central object in Boltzmann transport theory is the distribution matrix of the quasiparticles. In our case those are the Bogoliubov quasiparticles. We introduce a distribution function of quasiparticles of the form
\begin{eqnarray}
 f^i_{\sigma} (\bk,t) = \langle   \gamma^\dagger_{i  \sigma} (\bk,t)\gamma^{\phantom{\dagger}}_{i \sigma} (\bk,t)\rangle \;.
\end{eqnarray}

For all our following considerations it is important to assume that the Bogoliubov particles constitute reasonably sharp quasiparticles. 
In equilibrium, {\it i.e.} in the absence of external perturbations (such as an applied voltage, temperature gradient, ...), the distribution function is given by familiar Fermi-Dirac distribution
\begin{eqnarray}
f^i_{\sigma}(\bk,t)=n_f^0(\epsilon_i(\bk))=\frac{1}{e^\frac{ \epsilon_i(\bk)}{T}+1} \;,
\end{eqnarray}
where $\epsilon_i(\bk)$ is given in Eq.~\eqref{energies}. 

In order to deal with all the Bogoliubov quasiparticles on equal footing, we introduce a local basis, where the node is again parametrized by $i$. We assume that we apply a temperature gradient across the system, such that the temperature at position ${\bf{r}}$ is given by $T({\bf{r}})=T+{\bf{r}}\cdot {\bf{\nabla}}T$. 
The Boltzmann equation assumes the following schematic form
\begin{eqnarray}
\partial_t f^i (\bk,t)-X_i (\bk)=-\mathcal{I}^i_{\rm{coll}} (\bk)\;,
\end{eqnarray}
and since we are only interested in the time-independent solution we arrive at the following simple equation
\begin{eqnarray}\label{eq:boltzcompact}
X_i(\bk) =\mathcal{I}^i_{\rm{coll}}(\bk) \;.
\end{eqnarray}
We will now carefully disentangle the different terms in this expression.
The driving term $X_i$ assumes the generic form (we drop the spin-index for reasons of simplicity; it will trivially be accounted for by a factor of 2 in the end) 
\begin{eqnarray}\label{drive}
X_i(\bk)&=& \frac{\partial \epsilon_i(\bk)}{\partial \bk} \frac{{\bf{\nabla}}T}{T^2} \epsilon_i(\bk)n_f^0 (\epsilon_i(\bk)) (1-n_f^0 (\epsilon_i(\bk))) \;, \nonumber \\ \frac{\partial \epsilon_i (\bk)}{\partial \bk}&=& \frac{\left( \vfi (\vfi \cdot \bk)+\vdi (\vdi \cdot \bk)\right)}{\epsilon_i(\bk)} \;.
\end{eqnarray}
The form of Eq.~\eqref{drive} is markedly different from an isotropic system. In an isotropic system it is always possible to choose a basis such that one can formulate the problem in terms of the angle enclosed between $\nabla T$ and ${\bf k}$, {\it i.e.}, $X_i \propto v_F^2 |\nabla T| k \cos \left( \angle (\nabla T,{\bf k}) \right)$~\cite{FSMS,MFS}. This  allows to find a much simpler solution than in our case, where the angular dependence has to be taken seriously. Our strategy to properly account for the anisotropy will be to consider two independent equations for the particles moving parallel and perpendicular to the nodal points.

Furthermore $\mathcal{I}_{\rm coll}$ denotes the so-called collision term or integral. The collision integral $\mathcal{I}^i_{\rm{coll}}$ for a particle living at node $i$ will in general be composed of two sources of relaxation, namely the contribution due to disorder, henceforth called $\mathcal{I}^i_{\rm{dis}}$, and the contribution due to the inelastic scattering, $\mathcal{I}^i_{\rm{inel}}$, {\it i.e.},
\begin{eqnarray}
\mathcal{I}^i_{\rm{coll}}(\bk)=\mathcal{I}^i_{\rm{dis}}(\bk)+\mathcal{I}^i_{\rm{inel}}(\bk) \;.
\end{eqnarray}
The following sections are devoted to the understanding of the single and combined effect of those two scattering mechanisms.
The specific form of the two contributions is easily obtained and reads
\begin{widetext}
\begin{eqnarray}\label{eq:coll}
\mathcal{I}^i_{\rm dis}&=&2 \pi n_{\rm{imp}} \hat{u}^2 \sum_j \int \frac{d^2 k'}{(2\pi)^2}  \left (U^{\phantom{1}}_{i\bf{k}}\tau_z U^{-1}_{j\bf{k}'}\right)_{ii} \left (U^{\phantom{1}}_{j\bf{k}'}\tau_z U^{-1}_{i\bf{k}}\right)_{ii}\delta (\epsilon_i({\bf{k}})-\epsilon_j({\bf{k}}'))\left [f^{i}({\bf{k}})-f^{j}({\bf{k}}') \right]\nonumber\\ \mathcal{I}^i_{\rm inel}&=&- 2 \int d\omega' \int \frac{d^2 k'}{4\pi^2}\sum_{j} \left |\left(\hat{U}^{-1}_{i{\bf{k}}}\hat{U}^{\phantom{-1}}_{j{\bf{k}}+{\bf{k}'}}\right)_{ij}\right|^2 \delta \left(\omega+\omega'+\epsilon_j({\bf {k}}')\right) {\rm{Im}}D(\omega',{\bf{k}'}) \times \nonumber \\ &\times& \left [ f^i({\bf{k}})(1+f^j({\bf{k}}+{\bf{k}'}))+n({\bf{k}'})\left(f^i({\bf{k}}) -f^j ({\bf{k}}+{\bf{k}'}) \right) \right]
\end{eqnarray}
\end{widetext}
The set of equations Eq.~\eqref{heatcurrent},~\eqref{eq:boltzcompact},~\eqref{drive},~\eqref{eq:coll} is analyzed in the remainder of the paper. At this point we will perform a set of manipulations on the above expressions to bring them into a more tractable form.
It proves convenient to introduce the following simplifications at this point. Following Eq.~\eqref{localbasis} we can use the local basis at any nodal point, which allows to rescale momenta to yield an isotropic coordinate system, where the anisotropy is absorbed in the measure of the integral. We furthermore rescale momenta by a factor of $T$ which consequently eliminates all factors of $v_F$, $v_\Delta$, and $T$ except for global prefactors if scattering processes within one node or across the Brillouin zone are considered. If scattering processes mix adjacent nodes, however, this is not fully possible and factors of $\overline{v}$ prevail. The rescaling also allows the use of polar coordinates and implies that energies are now given by $\epsilon_i=\pm k T$. This allows to rewrite the scattering term locally as
\begin{eqnarray}
\tilde{X}_i&=&-\left( \vfi \xi_i+\vdi \Delta_i\right) k \frac{{\bf{\nabla}}T}{T} n^0_f (k) n^0_f (-k) \nonumber \\ &=&\tilde{X}_i^F+\tilde{X}_i^\Delta
\end{eqnarray}
where
\begin{eqnarray}\label{localbasis}
\xi_i(\theta)&=&=\cos\left(\theta+(i-1)\frac{\pi}{2}\right) \; ,\nonumber \\ \Delta_i(\theta)&=&\sin\left(\theta+(i-1)\frac{\pi}{2}\right)\;,
\end{eqnarray}
rendering all expressions dimensionless. The driving term splits into two parts, one characterizing parallel and one perpendicular electrons which is consistent with treatments using the Kubo formula~\cite{Durst}.

The structure of the driving term motivates an ansatz for the solution of the Boltzmann transport equation given by
\begin{eqnarray}\label{ansatz}
\delta \tilde{f}_i(k,\theta)&=&    {\bf{v}}_{F}^i \xi_i(\theta) \frac{{\bf v}_F^i{\bf{\nabla}}T}{T} n_f^0(k)n_f^0(-k) \Psi^i_F(k,\theta,\overline{v})\nonumber \\ &+&{\bf{v}}_{\Delta}^i \Delta_i(\theta) \frac{{\bf v}_\Delta^i{\bf{\nabla}}T}{T}n_f^0(k)n_f^0(-k) \Psi^i_\Delta(k,\theta,\overline{v}) \;,\nonumber \\
\end{eqnarray}
where the superscript $i$ of the functions $\Psi^i_F(k,\theta,\overline{v})$ and $\Psi^i_\Delta(k,\theta,\overline{v})$, respectively, accounts for the fact that we can define a local basis for every Dirac point itself. Again, it is worthwhile to contrast this expression from an isotropic system. In such a system we would not need to distinguish the two different nodal directions and we could simply choose an ansatz of the form $\Psi(|\bk|) \nabla T \cdot {\bf k} $ due to the spherical symmetry of the problem. 

A few words on the symmetry of the obtained expressions are in order here, since this will help to simplify life a lot in the following.The symmetry of the driving term under exchange of vis-a-vis nodes, {\it i.e.} $X_i (k,\theta)=X_{i+2}(k,\theta)$, enforces $\delta \tilde{f}_i(k,\theta)=\delta \tilde{f}_{i+2}(k,\theta)$. Furthermore we can deduce from the symmetries of the driving term that
\begin{eqnarray}\label{sympsi}
 \Psi^i_{F/\Delta}(k,\theta,\overline{v})= \Psi^i_{F/\Delta}(k,-\theta,\overline{v})= \Psi^i_{F/\Delta}(k,\theta\pm \pi,\overline{v})\;. \nonumber \\
\end{eqnarray}

Following Eq.~\eqref{heatcurrent} we can derive an expression for the energy current carried by the Bogoliubov particles of the form (note that we have $N_f=2$ in this expression)
\begin{eqnarray}\label{thermalcurrent}
{\bf{j}}_E&=&\frac{2T^2}{v_F v_\Delta}\sum_{i=1}^4 \int \frac{d\Omega_{\bf{k}} d k}{(2\pi)^2} k^2 \left[{\bf{v}}^i_F \xi_i + {\bf{v}}^i_\Delta \Delta_i \right] \delta \tilde{f}_i (k,\theta) \nonumber \\ &=&\frac{4T^2}{v_F v_\Delta}\sum_{i=1}^2 \int \frac{d\Omega_{\bf{k}} d k}{(2\pi)^2} k^2 \left[{\bf{v}}^i_F \xi_i + {\bf{v}}^i_\Delta \Delta_i \right] \delta \tilde{f}_i (k,\theta) \;.\nonumber \\
\end{eqnarray}
Taking into account all the aforementioned symmetries of the problem we finally arrive at a relatively simple expression for the thermal current carried by the quasiparticles under an applied thermal gradient across the sample 
\begin{eqnarray}
{\bf{j}}_E&=&{\bf V}_F \int \frac{d\Omega_{\bf{k}} d k}{(2\pi)^2} k^2  \xi_1^2(\theta)n_f^0(k)n_f^0(-k) \Psi^1_F(k,\theta,\overline{v}) \nonumber \\ &+&{\bf V}_\Delta \int \frac{d\Omega_{\bf{k}} d k}{(2\pi)^2}  k^2  \Delta_1^2(\theta)n_f^0(k)n_f^0(-k) \Psi^1_\Delta(k,\theta,\overline{v})    \;, \nonumber \\
\end{eqnarray}
where the terms mixing $v_F$ and $v_\Delta$ vanish according to Eq.~\eqref{sympsi}, due to
\begin{eqnarray}\label{symm}
\int d\theta \Delta_i(\theta) \xi_i(\theta)\Psi^i_{F/\Delta}(k,\theta,\overline{v})=0 \;.
\end{eqnarray}
This is analogous to the vanishing of mixed terms in the treatment by Durst and Lee~\cite{Durst}.
We furthermore introduced the following abbreviation:
\begin{eqnarray}
{\bf V}_{F/\Delta}=\frac{4T^2}{v_F v_\Delta} \left(  {\bf{v}}^1_{F/\Delta} {\bf{v}}^1_{F/\Delta} \nabla T+{\bf{v}}^2_{F/\Delta} {\bf{v}}^2_{F/\Delta} \nabla T  \right) \;.\nonumber \\
\end{eqnarray}
Since we are interested in the transport coefficient $\kappa_{xx}$ we can give its generic expression as
\begin{eqnarray}\label{kappa}
\kappa_{xx}&=&-4T^2\overline{v} \int \frac{d\Omega_{\bf{k}} d k}{(2\pi)^2}  k^2  \xi_1^2(\theta)n_f^0(k)n_f^0(-k) \Psi^1_F(k,\theta,\overline{v})   \nonumber \\&-&\frac{4T^2}{\overline{v}} \int \frac{d\Omega_{\bf{k}} d k}{(2\pi)^2}  k^2  \Delta_1^2(\theta)n_f^0(k)n_f^0(-k) \Psi^1_\Delta(k,\theta,\overline{v})  \;. \nonumber \\
\end{eqnarray}

\subsection{Thermal transport in a disordered $d$-wave superconductor}
\label{disorder}
Within this subsection we investigate the thermal transport in a system where the only relaxation mechanism is provided by dilute disorder. This has also been discussed in the thesis by Paaske~\cite{Paaske} and simply serves as a reference. We introduce a parameter measuring the strength of impurity scattering 
\begin{eqnarray}\label{alpha}
\alpha=\frac{2 n_{\rm{imp}}\hat{u}^2}{v_F v_\Delta} \;,
\end{eqnarray}
where
\begin{eqnarray}
\hat{u}^2=\left(V_{{\bf k}{\bf k}'}^{ij}\right)^2\;.
\end{eqnarray}
Expanding the collision term due to disorder to linear order in the deviation from equilibrium leaves us with the following expression
\begin{eqnarray}
\mathcal{I}^i_{\rm{dis}}(k,\theta)=\frac{\alpha k T }{4}    \int \frac{d \Omega_{\bf{k}'}}{2\pi}\sum_{j=1}^4T^+_{ij}(\theta,\phi)\left [\delta \tilde{f}_i (k,\theta)-\delta \tilde{f}_j(k,\phi) \right] \;,\nonumber \\
\end{eqnarray}
where we introduced the short-hand notation
\begin{eqnarray}
T^{\kappa}_{ij} (\theta,\phi)=1+\kappa \xi_i(\theta)\xi_j(\phi)-\kappa \Delta_i(\theta) \Delta_j(\phi) \;,
\end{eqnarray}
with $\xi_i(\theta)$ and $\Delta_i(\theta)$ defined in Eq.~\eqref{localbasis} and $\kappa=\pm$.
This factor simply accounts for the usual coherence factors~\cite{Schrieffer}. We are thus left with the task of solving the equation
\begin{eqnarray}
\tilde{X}_i(k,\theta)=\mathcal{I}^i_{\rm{dis}}(k,\theta) \;.
\end{eqnarray}
Plugging in the ansatz~\eqref{ansatz} we see that
\begin{eqnarray}
\mathcal{I}^i_{\rm{dis}}(k,\theta)&=&\alpha k T \delta \tilde{f}_i (k,\theta)-\frac{\alpha k}{2}    \sum_{j=1}^2\int \frac{d \Omega_{\bf{k}'}}{2\pi} \delta \tilde{f}_j(k,\phi)\nonumber \\ &=&\alpha k  T \delta \tilde{f}_i (k,\theta) \;.
\end{eqnarray}
where the second term in the first line vanishes due to the isotropic nature of the impurity scattering, which we assumed for reasons of simplicity. In the more generic case~\cite{Paaske} this is no longer true. In our problem, however, the second term will always vanish also in the presence of the inelastic scattering by virtue of Eq.~\eqref{sympsi}.

We can now find a simple solution given by
\begin{eqnarray}
\Psi^i_F(k,\theta,\overline{v})=\Psi^i_\Delta(k,\theta,\overline{v})=\frac{1}{\alpha} \;.
\end{eqnarray}
Using  Eq.~\eqref{kappa} it is straightforward to arrive at an expression for the thermal transport coefficient, given by
\begin{eqnarray}\label{thermaldisorder}
\kappa_{xx}=\frac{\pi T}{12 n_{imp}\hat{u}^2} \left (v_F^2+v_\Delta^2 \right)\;,
\end{eqnarray}
which is the central result of this section. Paaske~\cite{Paaske} showed that this discussion can easily be extended to include the case of anisotropic scattering, which means we are formally allowing for three different scattering matrix elements: intranodal scattering, which we denote $V_{\bf{k}{\bf{k}'}}^{11}$, scattering between adjacent nodes called $V_{\bf{k}{\bf{k}'}}^{12}$, and $V_{\bf{k}{\bf{k}'}}^{13}$ for scattering across the Brillouin zone. In this more general case the thermal conductivity reads
\begin{eqnarray}
\kappa_{xx}=\frac{\pi T}{12 n_{imp}u_0^2} \left (\frac{v_F^2}{1-\delta}+\frac{v_\Delta^2}{1+\delta} \right)\;,
\end{eqnarray}
where
\begin{eqnarray}
u_0^2&=&\frac{\left(V_{\bf{k}{\bf{k}'}}^{11}\right)^2+\left(V_{\bf{k}{\bf{k}'}}^{13}\right)^2+2\left(V_{\bf{k}{\bf{k}'}}^{12}\right)^2}{4}\;, \nonumber \\  \delta&=&\frac{\left(V_{\bf{k}{\bf{k}'}}^{11}\right)^2-\left(V_{\bf{k}{\bf{k}'}}^{13}\right)^2}{2u_0^2}\;.
\end{eqnarray}
This expression nicely reduces to Eq.~\eqref{thermaldisorder} for isotropic scattering, {\it i.e.}, $V_{\bf{k}{\bf{k}'}}^{11}=V_{\bf{k}{\bf{k}'}}^{12}=V_{\bf{k}{\bf{k}'}}^{13}$. 

In the following sections we will concentrate on isotropic scattering to study the interplay of disorder scattering and scattering from the nematic order parameter for reasons of simplified analysis. Another comment on this result is in order here. It has been shown that in the context of unconventional superconductor a faithful description of disorder requires to consider unitary scatterers, see Ref.~\cite{Gorkov}. This means that a finite concentration of non-magnetic impurities induces a finite density of states. The so-called universal conductivity obtains in the limit when temperature $T$ is much smaller than the impurity bandwidth $\gamma$, {\it i.e.} $T \ll \gamma$, see Ref.~\onlinecite{Graf}. The universal heat conductivity~\cite{Graf,Durst} has been shown to be independent of details of the disorder distribution, which means that the impurity concentration drops out in the final expression for the heat conductivity. Experimental evidence also points towards the existence of a universal conductivity in the limit $T \to 0$ independent of the scattering strength~\cite{Taillefer,Nakamae,Suzuki}.

Here, the result depends upon the impurity concentration explicitly. Our calculation, however, does not address the unitary limit but the Born limit, such that this discrepancy does not constitute a problem. In order to make contact with the unitary scattering limit, the above treatment must be (in the spirit of a quantum Boltzmann equation~\cite{Graf}) supplemented by a field renormalization stemming from the real part of the self-energy (this is neglected in the above Boltzmann equation, but can be incorporated in a straightforward manner~\cite{Rammer,Prange}), which cancels the explicit dependence upon the impurity scattering. This establishes the equivalence of the Kubo-formula calculations employing a self consistent Born approximation~\cite{Durst}, where the impurity scattering induces a finite density of states at the Fermi level.

\subsection{Thermal transport at the nematic transition}
\label{full}

Within this section we consider the electronic scattering off the nematic mode. The central approximation in this section is to assume the bosonic sector to be in equilibrium, which implies that the bosonic system is not excited by the temperature gradient and relaxes on a time scale faster than the time-scale associated with the electronic quasiparticles. One argument in favor of this point of view is to consider a situation in which the temperature gradient is applied along the $v_F$ direction of the electrons at the nodal point denoted 1. Under the formal assumption that the rest of the electronic quasiparticles is accelerated by the driving field in their $v_\Delta$ direction, they are in the limit of large anisotropy subdominant in their contribution to the thermal transport. The propagator of the bosonic mode
is linked to contributions from all the fermionic nodes, and so its response to the thermal gradient is suppressed by a
factor of $1/N$, where $N$ is the number of nodes. 
An analogous discussion has been carried out in the study of transport in bosonic theories in the large-N limit~\cite{SSbook}.

Another argument in favor of this approximation comes from rewriting the problem integrating out the bosons, which leads to a purely fermionic problem, see Appendix~\ref{zeromode}. In the fermionic language one can clearly see that a scattering involving fermions at adjacent nodes can relax the thermal current, whereas scattering between electrons at the same node or nodes across the Brillouin zone cannot. This argument, however, only works for large anisotropy, {\it i.e.} $\overline{v} \gg 1$. 

We first elaborate on the collision integral stemming from the scattering of the nodal fermions from the nematic order parameter fluctuations. It assumes the following generic form
\begin{widetext}
\begin{eqnarray}\label{scatterinint}
\mathcal{I}^1_{\rm{inel}}&=&\frac{T\lambda^2}{4 v_F v_\Delta N_f}  \int \frac{d\Omega_{\bf{k}'}}{2\pi}d k' k' T_{11}^-(\theta,\phi) D_{{\bf{k}}-{\bf{k}}'}^{''k-k'} \left( \left( n_B(k-k')+n_f^0(-k') \right)\delta \tilde{f}_1 (k,\theta) -\left( n_B(k-k')+n_f^0(k)\right) \delta \tilde{f}_1 (k',\phi) \right) \nonumber \\ &+&\frac{T\lambda^2}{4 v_F v_\Delta N_f}  \int \frac{d\Omega_{\bf{k}'}}{2\pi}d k' k' T_{13}^-(\theta,\phi) D_{{\bf{k}}-{\bf{k}}'}^{''k+k'} \left( \left( n_B(k+k')+n_f^0(k') \right)\delta \tilde{f}_1 (k,\theta) -\left( n_B(k+k')+n_f^0(k)\right) \delta \tilde{f}_1 (k',\phi) \right)\;,\nonumber \\
\end{eqnarray}
\end{widetext}
which is a generalization of the expressions shown in Refs.~\onlinecite{Hlubina,Rosch} accounting for the coherence factors and the different nodes. Furthermore, we exploit the fact that $\delta \tilde{f}_i(k,\theta)=\delta \tilde{f}_{i+2}(k,\theta)$ and introduced ${\tilde{D}}_{{\bf{k}}-{\bf{k}}'}^{''k-k'}$ being $\gamma$ times the imaginary part of the retarded Green's function of the bosonic modes, which was introduced in its imaginary frequency form in Eq.~\eqref{boseprop}. 
The full problem is analytically not tractable and has to be solved numerically. We thus use a variational approach~\cite{ziman,Hlubina,Rosch,Yaffe}, which allows to determine a bound for the conductivity. The interaction parameter $\lambda$ drops out of the problem exactly which simply reflects the fact that the present perturbation theory is not controlled in the smallness of $\lambda$, but in the smallness of $1/N_f$. It is interesting to note that the functions $\Psi_{F/\Delta}(k,\theta,\overline{v})$ acquire a true angular dependence in contrast to the pure isotropic impurity scattering problem. The presence of the nematic mode leads to a non-trivial renormalization of the velocity-parameter $\overline{v}$, see Ref.~\onlinecite{Yejin}, which will be taken into account later. 

We set up the variational problem defining the appropriate matrix elements 
\begin{widetext}
\begin{eqnarray}\label{functional}
{\cal Q}&=&{\cal Q}_F + {\cal Q}_\Delta={\cal X}-{\cal N}-{\cal D}\nonumber \\
{\cal X}&=&\frac{T}{{\bf{v}}_F^i \nabla T}\int \frac{k dk d \theta}{(2\pi)^2} \cos \theta \Psi^i_F(k,\theta,\overline{v}) \tilde{X}_i^F(k,\theta,\overline{v}) + \frac{T}{{\bf{v}}_\Delta^i \nabla T}\int \frac{k dk d \theta}{(2\pi)^2} \sin \theta \Psi^i_\Delta(k,\theta,\overline{v}) \tilde{X}_i^\Delta \nonumber \\ {\cal N}&=& \frac{4 T}{N_f} \int \frac{k dk d \theta}{(2\pi)^2} \frac{d\phi}{2\pi}d k' k' T^-_{ii}(\theta,\phi)\tilde{D}_{{\bf{k}}-{\bf{k}}'}^{''k-k'}n_f^0(k)n_f^0(-k) \left( n_B(k-k')+n_f^0(-k') \right)\left (\cos \theta \Psi_F^i(k,\theta,\overline{v})-\cos \phi \Psi_F^i(k',\phi,\overline{v}) \right)^2 \nonumber \\ &+&\frac{4 T}{N_f}  \int \frac{k dk d \theta}{(2\pi)^2} \frac{d\phi}{2\pi}d k' k' T^-_{ii}(\theta,\phi)\tilde{D}_{{\bf{k}}-{\bf{k}}'}^{''k-k'}n_f^0(k)n_f^0(-k) \left( n_B(k-k')+n_f^0(-k') \right)\left (\sin \theta \Psi_\Delta^i(k,\theta,\overline{v})-\sin \phi \Psi_\Delta^i(k',\phi,\overline{v}) \right)^2  \nonumber \\ &+& \frac{4 T}{N_f} \int \frac{k dk d \theta}{(2\pi)^2} \frac{d\phi}{2\pi}d k' k' T^-_{ii+2}(\theta,\phi)\tilde{D}_{{\bf{k}}-{\bf{k}}'}^{''k+k'}n_f^0(k)n_f^0(-k) \left( n_B(k+k')+n_f^0(k') \right)\left (\cos \theta \Psi_F^i(k,\theta,\overline{v})-\cos \phi \Psi_F^i(k',\phi,\overline{v}) \right)^2 \nonumber \\ &+&\frac{4 T}{N_f}  \int \frac{k dk d \theta}{(2\pi)^2} \frac{d\phi}{2\pi}d k' k' T^-_{ii+2}(\theta,\phi)\tilde{D}_{{\bf{k}}-{\bf{k}}'}^{''k+k'}n_f^0(k)n_f^0(-k) \left( n_B(k+k')+n_f^0(k') \right)\left (\sin \theta \Psi_\Delta^i(k,\theta,\overline{v})-\sin \phi \Psi_\Delta^i(k',\phi,\overline{v}) \right)^2 \nn \\ {\cal D}&=& \frac{\alpha T}{2} \int \frac{k dk d \theta}{(2\pi)^2} k n_f^0(k)n_f^0(-k) \left ( \cos \theta  \Psi_F^i(k,\theta,\overline{v})\right)^2 +\frac{\alpha T}{2} \int \frac{k dk d \theta}{(2\pi)^2} k n_f^0(k)n_f^0(-k) \left ( \sin \theta  \Psi_\Delta^i(k,\theta,\overline{v})\right)^2 \; ,
\end{eqnarray}
\end{widetext}
where ${\cal X}$ encodes the driving term, ${\cal N}$ denotes the scattering from the nematic mode, and ${\cal D}$ is simply the scattering from impurities. The above form is relatively simple and its  simplicity stems from the fact that there are a couple of different symmetries which can be exploited. In the following calculation we will assume $N_f=2$.
In this framework it is possible to obtain the Boltzmann transport equation by demanding a maximization of the functional in the following sense
\begin{eqnarray}
\frac{\partial {\cal Q}}{\partial \cos \theta \Psi_F^i(k,\theta,v)}=\frac{\partial {\cal Q}}{\partial \sin \theta \Psi_\Delta^i(k,\theta,v)} =0 \;.
\end{eqnarray}
By virtue of Eq.~\eqref{symm} the contribution associated with the direction perpendicular to the Fermi surface and the one associated with the parallel (gap) direction decouple nicely, and we are left with two equations that we can solve independently. Furthermore, comparing the expression for ${\cal X}$ in Eq.~\eqref{functional} and the expression for the transport coefficient $\kappa_{xx}$ given in Eq.~\eqref{kappa} we see that they are identical up to prefactors. This allows to determine $\kappa_{xx}$ knowing ${\cal X}$. In order to solve the above integral equation we can now make a variational ansatz and maximize the above functional with respect to a set of expansion coefficients, allowing to determine a lower bound on the contribution~\cite{ziman}. It seems sensible to choose the following ansatz
\begin{eqnarray}
\Psi_{F/\Delta}^i(k,\theta,\overline{v})=-a_{F/\Delta}(\overline{v})-b_{F/\Delta}(\overline{v}) \cos 2 \theta \;,
\end{eqnarray}
which again amounts to dealing with fermions perpendicular and parallel to the Fermi surface, separately. In the above expression $a_{F/\Delta}$ and $b_{F/\Delta}$ are the variational parameters, which have to be determined as to minimize the functional.

Of course, the whole series $\cos 2 n \theta$ with n being an integer is allowed for symmetry reasons. However, it is easy to check that only $n=1$ contributes to the heat current, which is why we concentrate on this mode. We choose the mode with no k-dependence, since this is the mode associated with energy conservation in a clean system, see Appendix~\ref{zeromode}. This mode in a clean system with a fully relativistic Hamiltonian is not relaxed and leads to an infinite thermal conductivity. One can of course improve upon the approximation by including more modes, which is however beyond the scope of this work, since we are mainly interested in qualitative features. It is important to note that the expressions for ${\cal X}$ and ${\cal D}$ can be calculated analytically, yielding
\begin{eqnarray}
{\cal X}&=& \frac{\pi}{24} \left (a_F+a_\Delta+\frac{ b_F}{2}+ \frac{b_\Delta}{2} \right) \nonumber \\ {\cal D}&=& \frac{\alpha T \pi}{48}\left ( a_F^2+a_\Delta^2 + \frac{b_F^2}{2}+ \frac{b_\Delta^2}{2}+a_F b_F + a_\Delta b_\Delta\right)\;. \nonumber \\
\end{eqnarray}
The thermal conductivity will finally be given by the simple expression
\begin{eqnarray}\label{thermalcon}
\frac{\kappa_{xx}}{T}&=&\frac{\pi}{6} \overline{v} \left[a_F(\overline{v}) +\frac{b_F(\overline{v})}{2} \right]\nonumber \\ &+& \frac{\pi}{6}\frac{1}{\overline{v}} \left[a_\Delta(\overline{v})+\frac{b_\Delta (\overline{v})}{2} \right] \;.
\end{eqnarray}
At this stage it is worthwhile stressing the fact that the anisotropy parameter scales at criticality like
\begin{eqnarray}\label{asymptoticflow}
\overline{v}(T)=\log \left( \frac{1}{T}\right)\;.
\end{eqnarray}
It is an interesting question to analyze the role of the "vertex"-corrections, which are encoded in the coefficients $a_\Delta$, $a_F$, $b_\Delta$, and $b_F$. Having a closer look at Eq.~\eqref{thermalcon} reveals that for large initial anisotropy the second term will be subdominant. Asymptotically the heat conductivity is thus given by
\begin{eqnarray}\label{thermalcon1}
\frac{\kappa_{xx}}{T}&=&\frac{\pi}{6} \overline{v} \left[a_F(\overline{v}) +\frac{b_F(\overline{v})}{2} \right] \;.
\end{eqnarray}
\begin{figure}[h]
\includegraphics[width=0.45\textwidth]{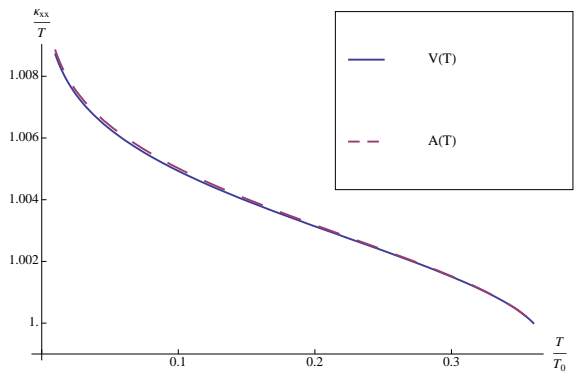}
\caption{(Color online) The uppermost curve (solid/blue) shows the temperature behavior of $V(T)=\frac{\overline{v}(T)}{\overline{v}(T_0)}$. The lower curve (short dashes/red) shows ratio $A(T)=\frac{a_\Delta(\overline{v}(T))+\frac{b_\Delta(\overline{v}(T))}{2}}{a_\Delta(\overline{v}(T_0))+\frac{b_\Delta(\overline{v}(T_0))}{2}}$ plotted with an initial value of $v_F/v_\Delta=20$ at a temperature $T_0$.}\label{fig:comparison}
\end{figure}
A closer inspection of the numerics reveals that the flow of the velocity anisotropy and the flow of the "vertex" (the expression in the brackets) have the same logarithmic dependence upon temperature. This is shown in Fig.~\ref{fig:comparison}, where $V(T)=\frac{\overline{v}(T)}{\overline{v}(T_0)}$ is compared to $A(T)=\frac{a_\Delta(\overline{v}(T))+\frac{b_\Delta(\overline{v}(T))}{2}}{a_\Delta(\overline{v}(T_0))+\frac{b_\Delta(\overline{v}(T_0))}{2}}$. This implies an asymptotic behavior of the heat conductivity according to
\begin{eqnarray}
\frac{\kappa_{xx}}{T} \propto \log^2 \left( \frac{1}{T}\right)\;.
\end{eqnarray}
The role of the vertex is also highlighted in the subsequent discussion and in Fig.~\ref{fig:vertex}.
\section{Thermal conductivity at the nematic phase transition}
We now turn our attention to the numerical solution of Eq.~\eqref{thermalcon} in various situations. 

In a first part, Sec.~\ref{thermalclean}, we will analyze a clean system, in which the only current relaxation stems from inelastic scattering of the electronic quasiparticles from the effective bosonic mode. In a second step, Sec.~\ref{thermalfull}, we will add dilute disorder to the problem and consider the full problem.

It was shown that the nematic phase transition is described by a fixed point located at infinite anisotropy, {\it i. e. } $v_F/v_\Delta \to  \infty$~\cite{Yejin} . This fixed point is approached in a logarithmic manner upon lowering the temperature. This implies that in order to do a realistic calculation of the thermal conductivity one has to take into account the logarithmic flow of the velocity-anisotropy, which will lead to a logarithmic enhancement of the thermal conductivity upon lowering the temperature. 

The RG flow equations of the velocities which need to be integrated to achieve this are reviewed in Appendix~\ref{RG}.

All the plots presented in this section originate from a combination of a numerical solution of Eq.~\eqref{thermalcon} and the flow equations presented in Appendix~\ref{RG}.

\subsection{Quasiparticle thermal conductivity in a clean system}\label{thermalclean}

\begin{figure}[h]
\includegraphics[width=0.45\textwidth]{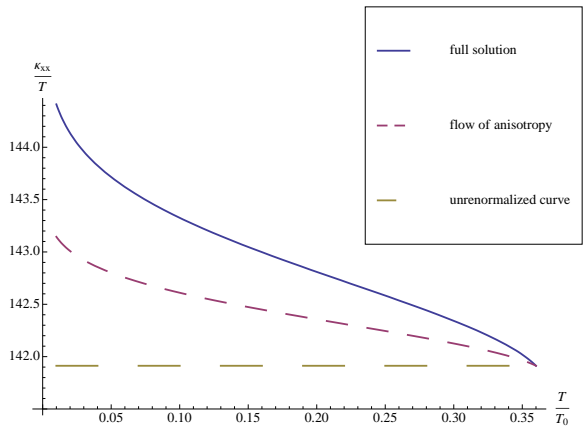}
\caption{(Color online) The uppermost curve (solid/blue) shows the full numerical solution of Eq.~\eqref{thermalcon}, taking into account the temperature renormalization effect of the anisotropy ratio. The middle curve (short dashes/red) shows the universal limit conductivity with running couplings, whereas the lowest curve (long dashes/green) simply serves as a reference curve. All plots employ an initial anisotropy ratio $v_F/v_\Delta=20$ at a temperature $T_0$.}\label{fig:vertex}
\end{figure}

In this section we consider a clean system, in which the only source of thermal current relaxation stems from inelastic scattering.
Experimental evidence suggests a value of $v_F/v_\Delta \approx 20$. Our reference point in the following plot is thus given by this ratio; since we are only interested in the qualitative features the initial temperature of the integrated flow is chosen arbitrarily and called $T_0$. From Eq.~\eqref{thermalcon} it is obvious, that there are different contributions, which we aim to disentangle. One is the obvious dependence on $\overline{v}$ and thus on its asymptotical flow (see Eq.~\eqref{asymptoticflow}). Another one comes from the implicit dependence of the parameters $a_{F,\Delta}$ and $a_{F,\Delta}$ on $\overline{v}$ and thus on $T$. In order to disentangle this, we have Fig.~\ref{fig:vertex} contain a comparison of three curves in the clean limit. The uppermost curve is a full numerical solution of Eq.~\eqref{thermalcon}, where all parameters and their respective RG flow are fully incorporated, {\it i.e.} the full dependence of $a,b$ on $\overline{v}$, and thus implicitly on $T$ is taken into account. The middle curve serves to contrast this from a situation, in which $a,b$ would only depend upon the initial value of $\overline{v}$, but their flow is not taken into account. This curve thus simply reflects the flow of the prefactors in Eq.~\eqref{thermalcon} alone and thus asymptotically mimics the flow of the velocity anisotropy alone, see Eq.~\eqref{asymptoticflow}. The third curve simple serves as a reference point showing an unrenormalized flat curve, denoting the value of the thermal conductivity obtained at $T_0$ with $v_F/v_\Delta = 20$.

In the next section we will additionally consider the interplay between disorder and interaction with the bosonic mode.

\subsection{Quasiparticle thermal conductivity in a disordered system}\label{thermalfull}

\begin{figure}[h]
\includegraphics[width=0.45\textwidth]{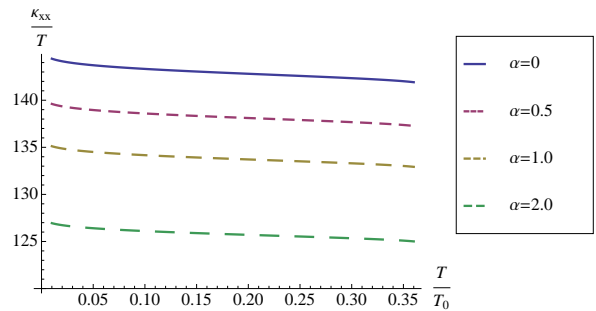}
\caption{(Color online) This plot shows different impurity strengths in increasing order from top to bottom with the uppermost curve being for the clean system ($\alpha=0$, $\alpha=0.05$, $\alpha=0.1$, $\alpha=0.2$), {\it i.e.} $\kappa_{xx}/T$, for an initial anisotropy ratio $v_F/v_\Delta=20$ at an arbitrary temperature $T_0$ for different disorder levels.}\label{fig:disorder}
\end{figure}

\begin{figure}[h]
\includegraphics[width=0.45\textwidth]{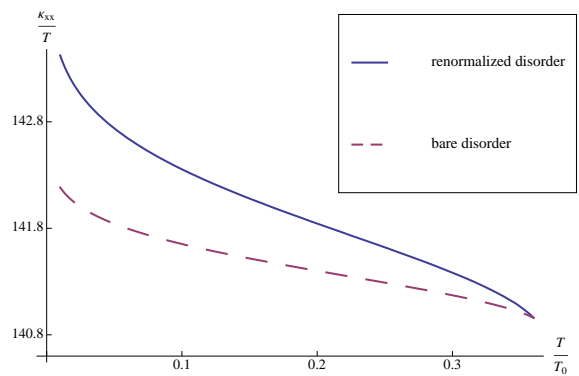}
\caption{(Color online) This plot shows the effect of disorder renormalization (initial value $\alpha=0.01$) due to the RG flow of the velocities. The upper curve (solid/blue) shows the full solution of Eq.~\eqref{thermalcon} taking into account the disorder renormalization, whereas the lower curve (dashed/red) is plotted at fixed disorder, but everything else renormalized. Again, {\it i.e.} $\kappa_{xx}/T$ is plotted for an initial anisotropy ratio $v_F/v_\Delta=20$ at an arbitrary temperature $T_0$.}\label{rundisorder}
\end{figure}

As we showed in the discussion of the thermal conductivity in a disordered $d$-wave superconductor, the impurity scattering strength does not vanish if the field renormalization is not taken into account, see Sec.~\ref{disorder}. This of course has consequences if one considers the problem taking into account both elastic and inelastic scattering, as shown in Fig.~\ref{fig:disorder}. We see that for different disorder levels, parametrized by $\alpha$, the curve has different offsets, being maximal in the clean system, as one would expect. 
On top of this, by the definition of $\alpha$ given in Eq.~\eqref{alpha}, it is obvious that the disorder strength is a flowing parameter under the renormalization group transformation, since it explicitely depends on the velocities. This effect is further illustrated in Fig.~\ref{rundisorder}, where full solutions of Eq.~\eqref{thermalcon} are plotted. The difference is that the upper curve takes into account the disorder renormalization, whereas the lower curve, which is plotted for reference, neglects this effect. 

\section{Conclusion}
\label{conclusion}

In this paper we have shown that at the nematic transition there is an enhancement of the thermal conductivity due to renormalization effects towards the fully anisotropic RG fixed point at $(v_\Delta/v_F)^*=0$. The central result of this paper is thus given by Eq.~\eqref{thermalcon} and shown in Fig.~\ref{fig:vertex}. We addressed this experimentally relevant issue in the framework of the Boltzmann equation for the fermions, arguing that thermal transport of the collective bosonic excitation is suppressed in the large 
$N_f$ limit \cite{SSbook}. Even though our approach strictly speaking is only justified under the conditions explained in Sec.~\ref{full}, we believe that our treatment correctly captures the logarithmic enhancement as the central quantitative signature of the nematic transition. The full treatment including the bosonic mode is beyond the scope of this paper and requires a full discussion in the framework of coupled Keldysh equations or a treatment in the framework of the Kubo formula or equivalently the memory matrices~\cite{Forster}. 

\acknowledgments We thank J. Paaske, M. M\"uller, Y. Huh, E.-A. Kim, M. Lawler, M. Vojta, and B. Wunsch for useful discussions and suggestions. This research was supported by the Deutsche Forschungsgemeinschaft under grant FR 2627/1-1 (LF), and by the NSF under grant DMR-0757145 (SS, LF).
\\

\appendix

\section{RG equations}\label{RG}
The full details of the derivation of the renormalization group equations were presented elsewhere~\cite{Yejin}, so we only repeat the crucial formulae. The methodology of the applied RG approach differs slightly from the more standard hard cutoff scheme, since a soft cutoff is more favorable in order to deal with the anisotropic velocities. Under the renormalization group transformation the fermionic velocities $v_F$ and $v_\Delta$ modify according to
\begin{equation}
\frac{d v_F} {d \ell}  = (C_1 - C_2) v_F \label{vfl}
\end{equation}
and
\begin{equation}
\frac{d v_\Delta}{d \ell} = (C_1 - C_3) v_\Delta . \label{vdl}
\end{equation}
This implies that the ratio of the two velocities scales like
\begin{equation}
\frac{d (v_\Delta/v_F)}{d \ell} = (C_2 - C_3) (v_\Delta/v_F)\;, \label{vdfl}
\end{equation}
where the functions $C_1$, $C_2$, and $C_3$ are given by
\begin{eqnarray}
C_1({\overline{v}}) &=&  \frac{2\overline{v}^{-1}}{\pi^3 N_f} \int_{-\infty}^\infty dx \int_0^{2 \pi} d \theta 
 f^{+--}(x,\theta,\overline{v}) \mathcal{G} (x, \theta)
\nonumber \\
C_2({\overline{v}}) &=&  \frac{2\overline{v}^{-1}}{\pi^3 N_f} \int_{-\infty}^\infty dx \int_0^{2 \pi} d \theta 
 f^{-+-}(x,\theta,\overline{v}) \mathcal{G} (x, \theta)
\nonumber \\
C_3({\overline{v}}) &=&  \frac{2\overline{v}^{-1}}{\pi^3 N_f} \int_{-\infty}^\infty dx \int_0^{2 \pi} d \theta 
f^{++-}(x,\theta,\overline{v}) \mathcal{G} (x, \theta)\;.
\nonumber \\
\label{rg5}
\end{eqnarray}
Furthermore, 
\begin{eqnarray}
\mathcal{G}^{-1} (x, \theta) &=& \frac{x^2 + \sin^2 \theta}{\sqrt{
x^2 + \overline{v}^{-2} \cos^2 \theta + \sin^2 \theta}} \nonumber \\&+&   \frac{x^2 + \cos^2 \theta}{\sqrt{
x^2 + \cos^2 \theta + \overline{v}^{-2} \sin^2 \theta}} \;
\end{eqnarray}
is the $\phi$ propagator inverse and
\begin{eqnarray}
f^{abc}(x,\theta,\overline{v})= \frac{(a x^2 +b \cos^2 \theta +c \overline{v}^{-2} \sin^2 \theta)}{(x^2 + \cos^2 \theta + \overline{v}^{-2} \sin^2 \theta)^2}
\end{eqnarray}
with $a,b,c=\pm$. This is the full set of equations required for the calculation of all running parameters used in Sec.~\ref{full}. 

\section{Contribution of the effective bosonic mode to the heat current}\label{bosoniccurrent}
The derivation of the heat current operator in interacting electronic systems is a long-standing problem stemming from the fact that a temperature gradient cannot be represented as a mechanical perturbation~\cite{Langer,Moreno,Catelani,Larkin}. Commonly, a Lagrangian approach is adopted to derive the appropriate relations. In the case of an electronic system interacting with a bosonic system, the microscopic expression for the heat current was derived for the case of phonons by Vilenkin {\it et al.}~
\cite{Vilenkin}. This procedure, however, does not exactly apply to our problem, due to the fact that the effective low-energy action of the bosonic mode is created by the electrons themselves. In general, for the system described by Eqs.~\eqref{$d$-wave}, \eqref{eq:inter}, and \eqref{eq:bos} it is practically impossible to disentangle the contributions to the energy current due to the electronic and the bosonic degrees of freedom. 
The full action of our problem at hand is given by the electronic $d$-wave superconductor coupled to the bosonic nematic mode, both of which can carry heat current. In a generic electron-phonon system the complete expression for the heat operator was given by Vilenkin {\it et al.}~\cite{Vilenkin}. Our problem, however, is different in the sense, that the effective dynamics of the bosonic mode is created by the electrons, see Eq.~\eqref{boseprop}. This is very similar to problems studied in the context of slave particle theories for the t-J model or in composite fermion theories of the fractional quantum hall effect, where an effective action for a U(1)-gauge field is generated and the analysis of transport quantities, especially in the d.c. limit, is tedious, see Refs.~\onlinecite{Ioffe,Nagaosa,YBKim1,YBKim2,Senthil}. 

\begin{figure}[h]
\includegraphics[width=0.45\textwidth]{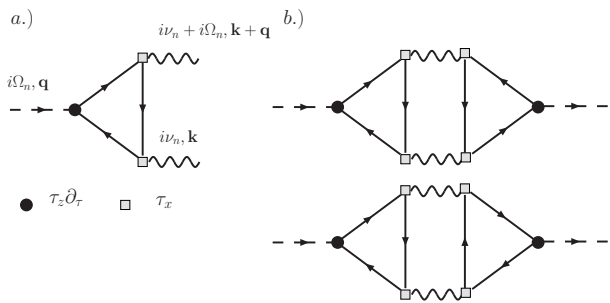}
\caption{the effective heat vertex proportional to $\overline{v}$ according to Eq.~\eqref{vertex} is shown in a.); b.) shows the bosonic contribution to the thermal current.}\label{fig:heatvertex}
\end{figure}

In order to derive an expression for the effective heat current carried by the bosonic mode we expand the saddle point action to third order following a minimal coupling scheme in the spirit of the minimal coupling to the electromagnetic field. The role of the electromagnetic field is in this case assumed by a thermal gauge field and is explained in great detail in Refs.~\onlinecite{Moreno,Larkin,Adrian}. In the following we assume a thermal gradient in the direction of the Fermi-velocity at node 1, without loss of generality. This implies we have a thermal gauge field, denoted $A_2^x$, without a y-component in the local basis defined at node 1. Furthermore, we will later make the approximation that $\overline{v} \gg 1$, which implies that the thermal gradient decouples from the second term in Eq.\eqref{$d$-wave}. We thus find the minimally coupled version of the problem to be given by
\begin{widetext}
\begin{eqnarray}
\mathcal{S}&=&\int d \tau \sum_{{\bf k},{\bf k}',\sigma} \Psi^\dagger_{1{\bf k}\sigma}\left ( \begin{array} {cc}  \partial_\tau- v_F (k_x \delta_{{\bf k},{\bf k}'}-i A_2^x({\bf{k}}-{\bf k}') i \overleftrightarrow{\partial}_\tau)& v_\Delta k_y \delta_{{\bf k},{\bf k}'}  \\  v_\Delta k_y \delta_{{\bf k},{\bf k}'} & \partial_\tau+v_F (k_x \delta_{{\bf k},{\bf k}'}-i A_2^x({\bf{k}}-{\bf k}') i \overleftrightarrow{\partial}_\tau)  \end{array} \right)   \Psi^{\phantom{\dagger}}_{1{\bf k}'\sigma} \nonumber \\ &+& \int d \tau \sum_{{\bf k},{\bf k}',\sigma} \Psi^\dagger_{2{\bf k}\sigma}\left ( \begin{array} {cc}  \partial_\tau- v_F k_y \delta_{{\bf k},{\bf k}'}& v_\Delta (k_x \delta_{{\bf k},{\bf k}'}-i A_2^x({\bf{k}}-{\bf k}') i \overleftrightarrow{\partial}_\tau) \\  v_\Delta (k_x \delta_{{\bf k},{\bf k}'}-i A_2^x({\bf{k}}-{\bf k}') i \overleftrightarrow{\partial}_\tau)  & \partial_\tau+v_F  k_y \delta_{{\bf k},{\bf k}'}  \end{array} \right)   \Psi^{\phantom{\dagger}}_{2{\bf k}'\sigma}\;,
\end{eqnarray} 
\end{widetext} 
where $\overleftrightarrow{\partial}_\tau=\frac{1}{2}\left( \overrightarrow{\partial}_\tau-\overleftarrow{\partial}_\tau\right)$, see Ref.~\onlinecite{Moreno}.
We can proceed with the derivation of the effective heat vertex of the bosonic mode, which is achieved calculating the diagram shown in Fig.~\ref{fig:heatvertex} a.). The "heat-vertex" with incoming frequency $\Omega_n$ and the external momentum ${\bf q}$ of the gauge field equal to zero, {\it i.e.} ${\bf q}={\bf 0}$, equates in the limit of zero temperature to 
\begin{widetext}
\begin{eqnarray}\label{vertex}
\Gamma (\nu_n,{\bf{k}},\Omega_n,{\bf q}={\bf 0})&=&\frac{\overline{v} N_f k_ x  }{16 \Omega_n }\left(\frac{v_F^2 k_x^2 + (\nu_n+\Omega_n)^2}{\sqrt{v_F^2 k_x^2 +v_\Delta^2 k_y^2+ (\nu_n+\Omega_n)^2}} - \frac{v_F^2 k_x^2+\nu_n^2}{\sqrt{v_F^2 k_x^2+v_\Delta^2 k_y^2+\nu_n^2}}\right) \nonumber \\ &+& \frac{N_f k_ x }{16 \Omega_n \overline{v}}\left( \frac{v_F^2 k_y^2 +\nu_n (\nu_n+\Omega_n) }{\sqrt{v_\Delta^2 k_x^2+v_F^2 k_y^2+\nu_n^2}}-\frac{v_F^2 k_y^2+\nu_n (\nu_n+\Omega_n) }{\sqrt{v_\Delta^2 k_x^2+v_F^2 k_y^2+(\nu_n+\Omega_n)^2}}\right)\nonumber \\ \lim_{\overline{v}\gg 1} &\approx&\frac{\overline{v} N_f k_ x  }{16 \Omega_n }\left( \sqrt{v_F^2 k_x^2 + (\nu_n+\Omega_n)^2}-\sqrt{v_F^2 k_x^2+\nu_n^2}\right) \;.
\end{eqnarray}
\end{widetext}
The derivation of this quantity is very tedious and involves the use of Feyman parameters, see {\it e.g.} Ref.~\onlinecite{Peskin}. However, it is interesting to note that the above expression, before taking the limit $1/\overline{v} \to 0$ in the last line, is {\it exact}. With the effective "heat-vertex" at hand we can proceed to calculate the bosonic contribution to the thermal transport.  The corresponding diagrammatic expressions assume the form shown in Fig.~\ref{fig:heatvertex} b.). At this stage it is very important to note that the above expression is even under $\nu_n \to -\nu_n$, $\Omega_n\to -\Omega_n$, and ${\bf k} \to -{\bf k}$. This is very important, since it implies that the heat-current-heat-current correlation function is given by 
\begin{eqnarray}
&&\mathcal{K}(\Omega_n,{\bf q}={\bf 0}) \propto \left(\overline{v} N_f\right)^2 T\sum_{\nu_n} \int \frac{d^2 {\bf k}}{(2\pi^2)}\nonumber \\ &&\Gamma (\nu_n,{\bf k},\Omega_n,{\bf 0})\left ( \Gamma (\nu_n,{\bf k},\Omega_n,{\bf 0})+ \Gamma (-\nu_n,-{\bf k},-\Omega_n,{\bf 0}) \right)  \nonumber \\   &&G_{\phi}({\bf k},\nu_n+\Omega_n)G_{\phi}({\bf k},\nu_n) \nonumber \\ &=& 2 \left(\overline{v} N_f\right)^2 T\sum_{\nu_n} \int \frac{d^2 {\bf k}}{(2\pi^2)}\nonumber \\ &&\Gamma (\nu_n,{\bf k},\Omega_n,{\bf 0})^2 G_{\phi}({\bf k},\nu_n+\Omega_n)G_{\phi}({\bf k},\nu_n) \;.
\end{eqnarray}
One comment has to be made at this stage, which is, that the heat-vertex in our case has been evaluated using the free Bogoliubov quasiparticle propagators. This, in general, constitutes a problem, once the d.c. limit in a response function is taken, since the resulting integrals are ill-defined and lead to infinite response coefficients~\cite{Mahan}. This is also reflected in the large-N calculation in a fermion-gauge field system as they often appear in gauge theory descriptions of strongly interacting electronic systems of Ref.~\onlinecite{YBKim1}, where the calculation becomes invalid in the low-frequency limit and the temperature has to serve as a cutoff. Starting from a heat-vertex calculated from free fermionic propagators, a calculation of the zero-frequency limit of the bosonic contribution to the thermal conductivity using the appropriate propagator for $\overline{v}\gg 1$ introduced in Eq.~\eqref{aniprop} following the prescription given in Ref.~\onlinecite{Adrian} does not yield a finite value. This can be traced back to the contribution of the free fermions and is an artifact of the neglect of the self-energies~\cite{Yamada}. We have not explicitly performed the calculation taking into account the self-energies. It is, however, straightforward using a spectral representation of the electronic contributions and introducing a thermal broadening~\cite{Aarts}. The important point is that the contribution of the bosonic modes to the thermal conductivity is finite and naively proportional to $\overline{v}^0$, which can easily be checked by taking the anisotropic bosonic propagator (Eq.~\eqref{aniprop}) and scaling all the momenta such as to make the resulting integral dimensionless. In this sense the bosonic contribution is down by $1/\overline{v}$ with respect to the fermionic contribution steming from the nodal fermions which move in perpendicular direction to the Fermi surface, compare Eq.~\eqref{kappa} in Sec.~\ref{Boltzmann}. However, it is unclear whether this power-counting argument applies once the self-energies are taken into account. This question is postponed to later works.
For a perfectly isotropic situation an analysis of the transport properties in terms of the fermionic model seems more appropriate, see Appendix~\ref{zeromode}, leading to an infinite thermal conductivity in the absence of disorder.

\section{Comments on finite conductivities in ideal systems}\label{zeromode}

Naively, in a clean system neglecting Umklapp scattering one would expect infinite response of the system to small perturbations, {\it i.e.} infinite transport coefficients, if interactions are momentum-conserving. 

If we consider a clean Fermi liquid with electron-electron interactions we observe the following. If we apply an electrical field across the system and forbid Umklapp scattering, no current is relaxed and the response is infinite. On the level of a Boltzmann equation, this can be seen very naturally. Assuming electronic quasi-particles, we can write the generic linear response form of the distribution function as
\begin{eqnarray}
f(\bk)=f^0(k)+f^0(k)(1-f^0(k))X_{\bf k}
\end{eqnarray}
Assuming a generic electron-electron interaction, we can write down the following schematic form~\cite{Mahan}
\begin{eqnarray}\label{scattint}
\mathcal{I}(\bk)=\int dk_1 dqF(k,k_1,q) \left[X_{\bf k}+X_{\bf k_1}-X_{\bf k-q}-X_{\bf k_1+q} \right]\;\nonumber \\ 
\end{eqnarray}
for the scattering integral.
It is obvious from the above equation that the right hand side vanishes for a particular choice of $X_{\bf{k}}$, namely
\begin{eqnarray}
X_{\bf k}=c {\bf k}\;,
\end{eqnarray}
which just restates the conservation of momentum. The associated mode is a so-called zero-mode of the scattering operator, which renders the matrix inversion singular. A zero mode is a mode, which cannot decay, {\it i.e.} the associated scattering time diverges, which also implies the conductivity as defined by the Drude~\cite{Mahan} formula to diverge.

One would expect the same kind of reasoning to apply for the thermal conductivity. However, the thermal conductivity usually is protected from this divergence (Refs.~\onlinecite{Mishchenko,Catelani,Moreno}). This can be related to the boundary condition of a vanishing electrical current, which forbids to excite the corresponding mode.

It is instructive to compare the above reasoning with a relativistically invariant (or equivalently strictly particle-hole symmetric) electronic theory, such as the Dirac theory in the way it applies to intrinsic graphene in the clean limit. It was pointed out in Refs.~\onlinecite{FSMS,MFS} that the conductivity in such a system can be finite, {\it i.e.} a momentum conserving interaction can relax a current due to the special particle-hole structure of the Dirac Hamiltonian. However, this is not true for the thermal conductivity, which is infinite. This can be traced back to the conservation of the energy component of the momentum-energy tensor and it is a generic property of relativistically invariant theories without any sort of translational symmetry breaking~\cite{Matthias,Kovtun}. Another way to state it is that in a relativistically invariant system the current and the heat current operator are orthogonal to each other, due to particle-hole symmetry, thus no boundary condition can cure the divergence.

This again can very nicely be seen in a Boltzmann transport approach. Here again, the collision integral assumes the generic form shown in Eq.~\eqref{scattint} and the existence of a zero-mode implies a diverging scattering time.

In our approach we reformulate a problem of interacting electrons in terms of electrons interacting with bosonic degrees of freedom. However, we will try to reformulate the present problem in terms of fermionic variables, yielding a faithful low-energy description. In order to do so we integrate out the effective bosonic degrees of freedom. This leads to a long-range electron-electron interaction, where the scattering operator for $\overline{v}=1$ looks identical to the scattering operator considered in a recent publication in the context of intrinsic graphene~\cite{FSMS,MFS}. In this case the thermal conductivity is infinite due to the presence of a momentum (energy) mode, which is not relaxed due to the presence of electron-electron interaction. We will show below that this changes in the case of anisotropic velocities. In order to do so we consider the ansatz introduced in Eq.~\eqref{ansatz}, where we chose a constant function $\Psi^i_F(k,\theta,\overline{v})=\Psi^i_\Delta(k,\theta,\overline{v})={\rm const}$. We can write the linearized version of the scattering integral as
\begin{widetext}
\begin{eqnarray}
\mathcal{I} &\propto&\sum_b \left [{\bf v}^a_F \nabla T ({\bf v}^a_F\cdot {\bf k}) +{\bf v}^b_F \nabla T ({\bf v}^b_F\cdot {\bf k}') -{\bf v}^a_F \nabla T ({\bf v}^a_F\cdot ({\bf k}+{\bf q}))  -{\bf v}^b_F \nabla T ({\bf v}^b_F\cdot ({\bf k}'-{\bf q})) \right. \nonumber \\  &+&\left.  {\bf v}^a_\Delta \nabla T ({\bf v}^a_\Delta\cdot {\bf k}) +{\bf v}^b_\Delta \nabla T ({\bf v}^b_\Delta\cdot {\bf k}') -{\bf v}^a_\Delta \nabla T ({\bf v}^a_\Delta\cdot ({\bf k}+{\bf q}))  -{\bf v}^b_\Delta \nabla T ({\bf v}^b_\Delta\cdot ({\bf k}'-{\bf q}))\right]\;.
\end{eqnarray}
\end{widetext}
In the case of no anisotropy, this expression vanishes identically under local rescaling of the corresponding variables. However, if $\overline{v}\neq 1$ this chances. The contribution from the scattering of electrons from one node to the same node, {\it i.e.} $a$ to $a$ or $b$ to $b$ can always be annihilated. However, scattering between adjacent nodes cannot, as can readily be seen from analyzing the expression corresponding to such an event
\begin{eqnarray}
\mathcal{I} &\propto &{\bf v}^a_F \nabla T ({\bf v}^a_F\cdot {\bf q})  \left ( \overline{v}^{-2}-1\right) \nonumber \\ &+& {\bf v}^a_\Delta \nabla T ({\bf v}^a_\Delta\cdot {\bf q})  \left ( \overline{v}^2-1\right) \;.
\end{eqnarray}
This implies that the energy mode is a zero mode of the problem for the isotropic system, {\it i.e.} $\overline{v}=1$, whereas the mode becomes massive once $\overline{v}^2\neq1$. This statement is equivalent to saying that the interaction with the nematic mode breaks the relativistic invariance of the underlying theory, which allows to have non-infinite thermal conductivity, since the divergence is not protected by symmetry any more. We thus expect the scattering of electrons from bosonic degrees of freedom (Eq.~\eqref{scatterinint}) to correctly describe this scattering for the case of extreme anisotropy, {\it i.e.} $\overline{v}\gg 1$, since then the bosonic theory seems to be the natural choice. The analysis of a clean electronic system and the interaction mediated thermal conductivity was carried out in Refs.~\onlinecite{Abrikosov,Mishchenko} and redone in a decoupling scheme in bosonic fields by Catelani and Aleiner~\cite{Catelani}.

\end{document}